\documentclass[lettersize,journal]{IEEEtran}
\IEEEoverridecommandlockouts
\usepackage{cite}
\usepackage{amsmath,amssymb,amsfonts}
\usepackage{algorithmic}
\usepackage{graphicx}
\usepackage{textcomp}
\usepackage{xcolor}
\usepackage{stfloats}
\usepackage{lettrine}
\usepackage[caption=false,font=normalsize,labelfont=sf,textfont=sf]{subfig}

\def\BibTeX{{\rm B\kern-.05em{\sc i\kern-.025em b}\kern-.08em
    T\kern-.1667em\lower.7ex\hbox{E}\kern-.125emX}}
\begin{document}

\title{BER Analysis of SCMA-OFDM Systems  in the Presence of Carrier
Frequency Offset\\

}

\author{Haibo Liu, Qu Luo,~\IEEEmembership{Graduate Student Member,~IEEE,}
        Zilong Liu,~\IEEEmembership{Senior Member,~IEEE,}\\
          Shan Luo,  
        Pei Xiao,~\IEEEmembership{Senior Member,~IEEE,}
        and Rongping Lin 
\thanks{

Haibo Liu and Shan Luo are with the School of Aeronautics and Astronautics, University of Electronic Science and Technology of China (UESTC), Chengdu 611731, China (e-mail: liu\_hb@std.uestc.edu.cn, luoshan@uestc.edu.cn).  Qu  Luo  and  Pei  Xiao  are  with  5G \& 6G  Innovation Centre, University of Surrey, UK (email: \{q.u.luo,   p.xiao\}@surrey.ac.uk). Zilong   Liu   is   with   the   School   of   Computer   Science   and   Electronics   Engineering,   University   of   Essex,   UK (email:   zilong.liu@essex.ac.uk). Rongping Lin is with the School of Information and Communication Engineering, UESTC, Chengdu 611731, China (e-mail: linrp@uestc.edu.cn).}
 }

\maketitle

\begin{abstract}
Sparse code multiple access (SCMA) building upon orthogonal frequency division multiplexing (OFDM) is a promising wireless technology for supporting massive connectivity in future machine-type communication networks. However, the sensitivity of OFDM to carrier frequency offset (CFO) poses a major challenge because it leads to orthogonality loss and incurs intercarrier interference (ICI).
In this paper, we   investigate the  bit error rate (BER) performance of SCMA-OFDM systems in the presence of CFO over both Gaussian and multipath Rayleigh fading channels. We first model the  ICI in SCMA-OFDM  as Gaussian variables conditioned on a single channel realization for fading channels.   The BER is then evaluated by averaging over all codeword pairs considering the fading statistics. Through simulations, we validate the accuracy of our BER analysis and reveal that there is a significant BER degradation for SCMA-OFDM systems when the normalized CFO exceeds 0.02.

\end{abstract}

\begin{IEEEkeywords}
Sparse code multiple access (SCMA), orthogonal frequency division multiplexing (OFDM), carrier frequency offset (CFO),  bit error rate (BER) 
\end{IEEEkeywords}

\section{Introduction}
Non-orthogonal multiple access (NOMA) has been extensively studied in recent  years to  meet the stringent requirements such as massive connectivity and higher spectral efficiency in future machine-type communication networks \cite{SparseLiu}.  In a NOMA system, multiple users communicate simultaneously to achieve an overloading  factor larger than one \cite{ChenMIMO,oNOMA}. As an important variant of code division multiple access, sparse code multiple access (SCMA) has emerged as a promising NOMA technique \cite{LuoLPSCMA}. In SCMA,  the  incoming message bits of each SCMA user are directly mapped to a multi-dimensional sparse codeword drawn from a carefully designed codebook \cite{tuofeng}. 
To deal with asynchronous wireless transmissions, one can build SCMA upon orthogonal frequency division multiplexing (OFDM) \cite{Prototypescma,wen2017multiple,li2022composite}, where the resultant system is called SCMA-OFDM. This is because the cyclic prefix (CP) in an OFDM system can help accommodate the inter-user asynchrony and intersymbol interference in uplink channels. In SCMA-OFDM,  multi-dimensional sparse codewords are sent over a number of orthogonal subcarriers (SCs),  also known as resource elements (REs). 

 Due to its multicarrier feature, nonetheless, an OFDM system may suffer from significant inter-carrier interference (ICI) due to the carrier frequency offset (CFO) caused by inaccurate oscillator clocks. As a matter of fact,  CFO may be large and widely present for an array of low-cost and low-end communication sensor devices.  In addition, similar to the CFO effect, the deleterious Doppler shifts in moving environments may also lead to ICI.  Extensive research activities have been conducted in recent years to understand the detrimental effects of CFO in OFDM systems \cite{Keller,Rugini,Armstrong,Dharmawansa}.  For example, in \cite{Keller}, the  ICI  was approximated to be a Gaussian distributed noise.   In \cite{Rugini}, the authors extended the approximation method of \cite{Keller} to frequency-selective channels in order to derive an improved bit error rate (BER) analysis framework.  
It should be noted that these works generally assume that the subcarriers in an OFDM symbol are independent.   However, this may not  held in SCMA-OFDM because   CFO simultaneously affects those nonzero codeword elements located in a frequency sub-band. Aiming to address this problem, we  carry out a systematic study to analyze the impact of CFO on SCMA-OFDM.

Despite a rich body of literature on SCMA, there is a paucity of  studies on its accurate error rate  performance analysis. The major difficulty lies in  the multidimensional sparse codebook nature of SCMA that involves  multi-user   and multi-carrier 
transmissions.   The error performances of uplink and downlink SCMA over     Rayleigh fading channels with a maximum likelihood   decoder were analyzed in \cite{BaoJoint}, and   a similar study was presented in \cite{LimUplink} for uplink SCMA with multiple receive antennas.

In this letter, we present a BER analysis for the   CFO impaired  SCMA-OFDM systems over  Gaussian and multipath Rayleigh fading channels.  First, the ICI in SCMA-OFDM is approximated by two uncorrelated Gaussian variables and the signal-to-interference-plus-noise ratio (SINR)    conditioned on a single channel realization is evaluated. Then,  the conditional pair-wise error probability (PEP) is obtained by using the conditional SINR.  Finally, the  BER is calculated by averaging over the     fading statistics and all the transmitted codewords.    Our analysis shows that the PEP of SCMA systems affected by CFO can be expressed as the product of a few integrals, where  each integral can be expressed by a series of generalized hyper-geometric functions. Interestingly, our numerical evaluations indicate that the BER performance of SCMA-OFDM system degrades significantly when the normalized CFO exceeds 0.02, thus providing a useful guideline for practical system design\footnote{\textit{Notations:} $x,\mathbf{x}$ and $\mathbf{X}$ denote scalar, vector and matrix, respectively.
$\text{diag}(\mathbf{x})$ represents a diagonal matrix with the main diagonal vector $\mathbf{x}$.
$\mathbf{X}^{\mathcal T} $ and $\mathbf{X}^{\mathcal H}$ denote the transpose and the Hermitian transpose, respectively.
${E}\{\cdot\}$ denotes the expectation operator.
$\mathbb{B}$ and $\mathbb{C}$ denote the set of binary numbers and complex numbers, respectively.
 }.

\section{System Model}

\subsection{SCMA Communication Model}
\begin{figure}[!htbp]
\centerline{\includegraphics[width=7.5cm]{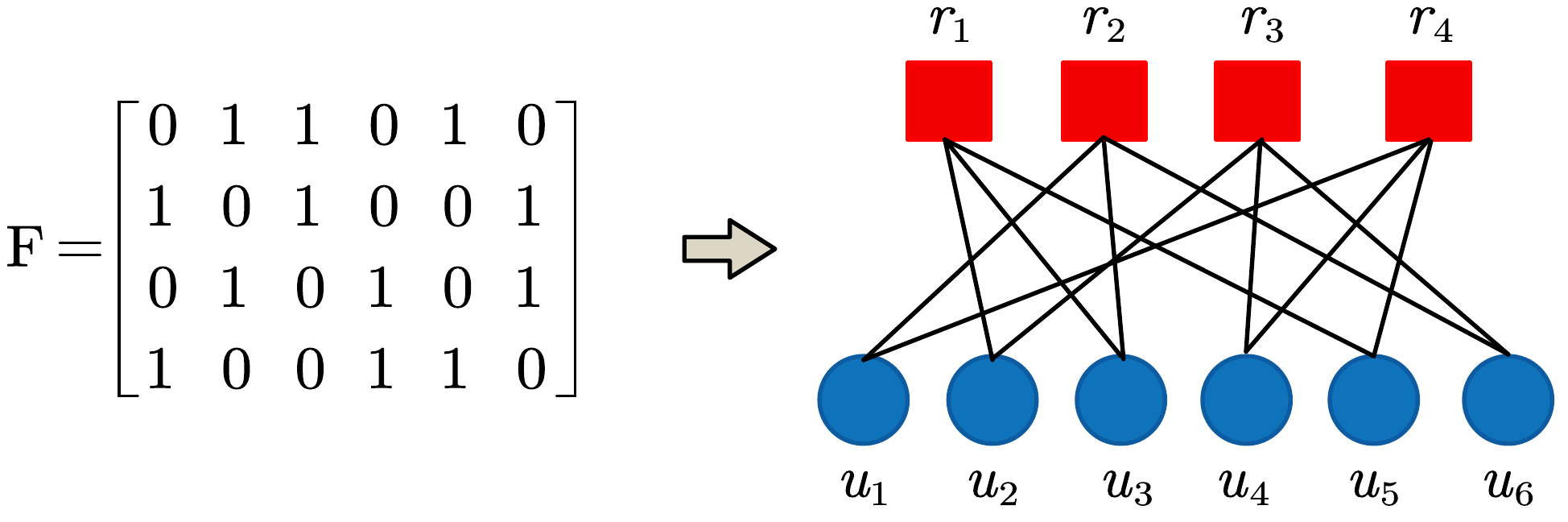}}
\caption{Factor graph representation for an SCMA system with $J = 6, K = 4,  V=2$.}
\label{facotGraph}
\end{figure}

We consider a downlink SCMA system, where $J$ users communicate over  $K$ resource elements, where $J>K$. The overloading factor defined as $J/K$ is thus larger than one. In SCMA, each user is assigned with  an unique codebook, denoted by  $ \boldsymbol {\mathcal {X}} _{j}=\left\{\mathbf{x}_{j, 1}, \mathbf{x}_{j, 2}, \ldots, \mathbf{x}_{j, M}\right\}   \in \mathbb {C}^{K \times M}, j \in\{1,2, \ldots, J\}$, consisting of $M$ codewords with a dimension of $K$.  During transmissions,  each user   maps $\log_2\left(M\right)$   binary bits to a length-$K$ codeword $ \mathbf {x} _{j}$ drawn from the $ \boldsymbol {\mathcal {X}}_{j}$.  The mapping process can be expressed as \cite{MultitaskSCMA}
\begin{equation}
\small
    f_{j}: \mathbb{B}^{\log _{2} M \times 1} \rightarrow \boldsymbol{\mathcal { X }}_{j} \in \mathbb{C}^{K \times M}, \text { i.e., } \mathbf{x}_{j}=f_{j}\left(\mathbf{b}_{j}\right),
\end{equation}
where $\mathbf{b}_{j}=\left[b_{j, 1}, b_{j, 2}, \ldots, b_{j, \log _{2} M}\right]^{\mathcal {T}} \in \mathbb{B}^{\log _{2} M \times 1}$ represents the input binary message  of the $j$th user. The $K$-dimensional complex codewords in the SCMA codebook are sparse vectors with $V$ non-zero elements and $V< K$. 
The factor graph  can be used to represent the connections between user nodes (UNs) and resource nodes (RNs) in SCMA. The sparse structure of the $J$ SCMA codebooks can also be represented by the indicator  matrix   $\mathbf {F}   \in \mathbb {B}^{K\times J}$.  An element of  ${\bf{F}}$ is defined as ${f_{k,j}}$ which takes the value of $1$ if and only if $j$th UN  ($u_j$) is connected to  $k$th RN  ($r_k$) and 0 otherwise.   Fig. \ref{facotGraph}  illustrates an SCMA indicator matrix and corresponding factor graph with $J=6$, $K=4$ and $V=2$.

\begin{figure*}[htbp!]
\centering
\includegraphics[width=0.8\textwidth]{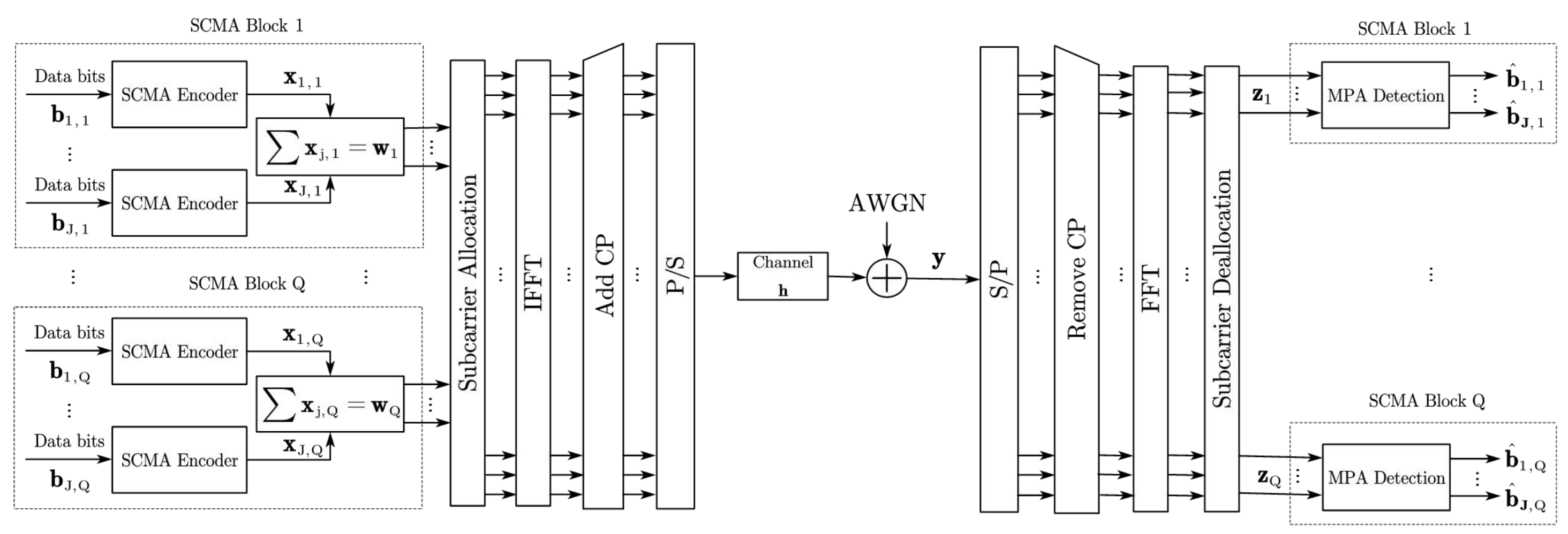} 
\caption{Downlink SCMA system based on OFDM.}
\label{diag_combination}
\vspace{-1em}
\end{figure*}

 \subsection{Downlink SCMA-OFDM  Impaired by CFO}
  In SCMA-OFDM systems,  the SCMA multiplexing  symbols are transmitted over multiple OFDM  SCs.  The
block diagram of a downlink SCMA system is shown in Fig. \ref{diag_combination}, where $N$  represents the number  of SCs in an OFDM system. We assume that $N= QK$, where $Q$ is the total number of SCMA blocks, each consisting of $K$ REs. In each  SCMA  block, users' data are first superimposed over $K$  REs, and then allocated to OFDM SCs. The superimposed  codewords at the $q$th SCMA block is denoted as $ \mathbf{w}_q=\sum_{j=1}^{J} \mathbf{x}_{j,q} \in \mathbb C^{K \times 1}.$ We introduce two  allocation schemes, namely the localized and interleaved allocations. 

In the case of localized transmission,  the superimposed codewords are mapped to a number of consecutive OFDM SCs. Therefore, the resultant signal is obtained as $\mathbf{s} = \left[\mathbf{w}_1^{\mathcal T}, \mathbf{w}_2^{\mathcal T}, \ldots, \mathbf{w}_Q^{\mathcal T}  \right]^{\mathcal T}$.

In the case of interleaved transmission, the $N$ SCs are divided into $K$ groups, with each group consists of $Q$ SCs. Then, the $k$-th entry of $\mathbf{x}_q$ is transmitted at the $q$-th position of group $k$ over the OFDM symbol. As a result, the mapped signal is obtained as $\mathbf{s} = \left[\mathbf{r}_{1}^{\mathcal T}, \mathbf{r}_{2}^{\mathcal T}, \ldots, \mathbf{r}_{K}^{\mathcal T}  \right]^{\mathcal T},$

  where $\mathbf{r}_{k} = \left[ {w}_{k,1}, {w}_{k,2}, \ldots, {w}_{k,Q} \right]$, and ${w}_{k,q}$ denotes the $k$th entry of $\mathbf{w}_q$. 

Afterwards, the transmitted signal in the time domain is obtained by performing the inverse fast Fourier transform (IFFT)  and adding a CP.  Let  $\mathbf{h}=[h_0,h_1, \cdots h_{P-1}]^{\mathcal {T}}$ be the multipath  fading channels   of $P$ paths. We  assume that the CP length is not less than $P$.    At the receiver side, in the presence of {CFO}\footnote{Similar to \cite{Keller}, we assume the CFO is the same for each path.},  the received   signal after down-sampling and  removing the CP can be expressed in   matrix form as 
\begin{equation}
\small
\label{NoCP}
\mathbf{y}=  \mathbf{D H} \mathbf{F}^{\mathrm{H}} \mathbf{s}+\mathbf {n},
\end{equation}
where  $\mathbf{D}   $ is an $N\times N$ diagonal matrix defined by $[\mathbf{D}]_{n, n}=\mathrm{e}^{j 2 \pi \varepsilon(n-1) / N}$, $\varepsilon$ is the  CFO normalized to the subcarrier spacing,   $\mathbf{H}$ is the circular channel matrix given by $[\mathbf H]_{n,m}= h_{(n-m)\mod N}, h_i=0, P \leq i  \leq N-1$, and $\mathbf {n}$ is the noise vector with the Gaussian noise with variance of $\sigma_0^2$.  After performing
FFT at the receiver, (\ref{NoCP}) can be expressed as
\begin{equation}
{\mathbf{z}}= \mathbf\Phi \mathbf\Lambda \mathbf{s}+{\mathbf{v}},
\end{equation}
where $\mathbf\Phi=\mathbf{F D F}^{\mathrm{H}}$ denotes the circular matrix that yields the ICI,    ${\mathbf{v}}=\mathbf{F} \mathbf{w}$ represents the  noise term, and $\boldsymbol{\Lambda}=\mathbf{F} \mathbf{H} \mathbf{F}^{\mathrm{H}}=\operatorname{diag}(\boldsymbol{\lambda})$ is the channel diagonal matrix with elements expressed by
\begin{equation}
\small
\label{freqH}
\boldsymbol{\lambda}=\sqrt{N} \mathbf{F h},
\end{equation}
where $\boldsymbol{\lambda}=[\lambda_{1} , \lambda_{2}, \ldots, \lambda_{N}]^{\mathrm{T}}$ denotes the channel vector in frequency domain. With these definitions, it is straightforward to verify that
\begin{equation}
\small
\mathbf{[\Phi]}_{n, m}=\frac{\sin (\pi((m-n) _{\text {mod } N}+\varepsilon))}{N \sin \left(\frac{\pi}{N}\left((m-n)_{_{\text {mod } N}}+\varepsilon\right)\right)}  \mathrm{e}^{j \pi \frac{N-1}{N}\left((m-n)_{\text {mod } N}+\varepsilon\right)}.
\end{equation}
Obviously, the matrix $\mathbf{\Phi}$ includes a phase-shift term $\mathrm{e}^{j \pi \frac{N-1}{N}\varepsilon}$ that applies to all SCs. In this paper, we assume perfect channel state information at the receiver.

\section{Performance Analysis of SCMA-OFDM in the presence of CFO}

In this section, we consider the interleaved transmission scheme  due to its larger frequency diversity.   
Let $\xi_N={\{1,2,\ldots,N\}}$ be the set of OFDM SCs. 
After the SC de-allocation, the received  $q$th SCMA block is denoted by   $\mathbf z_q =\{z_{q,k}\}  \in \mathbb C^{K \times 1}  $. 
In the following, we use the subscripts $k$ and $n$ for  SCMA REs  and  OFDM SCs, respectively.  For interleaved transmission, we have  $n = Q(k-1)+q$.
In order to evaluate the error probability and without loss of generality, we focus on the first SCMA block in the received OFDM symbol and drop the block index $q$ in $\mathbf{w}_q$ for the sake of simplicity.   Therefore,  ${z}_{k },  1 \leq k \leq K$  can be expressed as
\begin{equation}
\small
\label{rk}
\begin{aligned}
{z}_{k}=\phi  _{n,n }\lambda_{{n}} w_{{k}}+ \underbrace{\sum_{ m\in \xi_{N \backslash {n}} }  \phi  _{{n},m} \lambda_{m} s_{m}}_{\text{ICI}_{k}}  +v_{{n}},  \text{for  }  n= \text{Ind}_k, 
\end{aligned}
\end{equation}
where  $\text{Ind}_k = Q(k-1)+1$,  $\xi_{N \backslash {n}} = \xi_{N} {\setminus \{{n}\} }$ denotes the REs after removing the $n$th RE in $ \xi_{N \backslash {n}}$,  $w_k \in \mathbf w  $ is the $k$th transmitted codeword over the $n$th OFDM SC, and
$\phi _{n,m}=[\mathbf{\Phi}]_{n,m}$  represents the ICI coefficient from   subcarriers $\xi_{N \backslash n}$  to the subcarrier $n$. In the rest of the paper, unless otherwise stated,  $n= \text{Ind}_k$.   Assume that the energy of the transmitted codeword is normalized to unit.   For Gaussian channels, $\text{ICI}_{k}$  can be approximated as a Gaussian distributed random variable  with zero mean and  variance given by \cite{PIMRC}
\begin{equation}
\small
\label{ICI}
\begin{aligned}
  E\{\vert \text{ICI}_{k}\vert ^{2}\}  =  
&  \frac{J}{K} \Big (1- E\Big \{\vert\sum_{ m\in \xi_{N \backslash {n}} }  \phi  _{{n},m} \vert ^{2} \Big \} \Big)   \\ =
& \frac{J}{K}\left(1-{\sin^{2}\pi\varepsilon\over N^{2}\sin^{2}{\pi\varepsilon\over N}}\right).
\end{aligned}
\end{equation}
 In general, (\ref{ICI}) is tightly held as $N$ increases.  In the following, we present  the PEP in the SCMA-OFDM system  and the    ICI approximation for fading channels. Due to CFO and additive white Gaussian noise, the transmitted signal $\mathbf w$ is assumed to be erroneously decoded to another codewords $\hat{\mathbf{w}}$.  Denote  the   
 PEP between ${\mathbf{w}} $ and  $\hat{\mathbf{w}} $ as $\text{Pr} \{\mathbf{w} \to \mathbf{\hat{w}} \vert \boldsymbol { \lambda}\}$, which is  conditioned on  the  channel fading vector ${\boldsymbol \lambda}$, then  the average BER for SCMA-OFDM systems with  a maximum likelihood (ML) detector can be approximated as \cite{LuoLPSCMA} 
  \begin{equation}
  \label{ABER}
  \small
  \begin{aligned} 
  {\text {BER}} \leq &\frac {1}{M^{J} J\log _{2}(M)}  \\
  & \quad \sum _{\mathbf {w}} 
  \sum _{\hat {\mathbf {w}}\neq \mathbf {w}}{n_{\text {E}}(\mathbf {w},\hat {\mathbf {w}})}  \int\limits_{{\boldsymbol \lambda}} \text{Pr} \{\mathbf{w} \to \mathbf{\hat{w}} \vert \boldsymbol {\lambda}\}f_{\Lambda}({\boldsymbol \lambda}) d{\boldsymbol \lambda}, \\
  \end{aligned}
    \end{equation}
where   $n_e\left(\mathbf {w},\hat {\mathbf {w}}\right)$  denotes number of the erroneous bits when $\hat {\mathbf {w}}$ is decoded. It is noted that the evaluation of (\ref{ABER}) needs to deal with the  $ N$-dimensional probability density function (pdf)  $f_{\Lambda}({\boldsymbol \lambda})$.  As  the frequency domain channel is obtained by combining $P$ random variables, the variables in ${\boldsymbol \lambda}$ are correlated with one another. Consequently,  the  pdf  $f_{\Lambda}({\boldsymbol \lambda})$ cannot be factorized as the product of $N$ independent one-dimensional pdfs.  To address this problem, we re-write the multidimensional integration  as $f_{\Lambda}({\boldsymbol \lambda}) = f_{{{\Lambda}}_{n} \vert\lambda_{N \backslash {n}}}({ {\boldsymbol \lambda}_{N \backslash {n}}}\vert\lambda_{n}) f_{{ \lambda}_{n}}(\lambda_{n})   $, where   ${\boldsymbol\lambda}_{N\backslash n }$ denotes the channel vector ${\boldsymbol\lambda}$ after  removing the $n$th entry 
 and $f_{{{\Lambda}}_{n} \vert\lambda_{N \backslash {n}}}({ {\boldsymbol \lambda}_{N \backslash {n}}}\vert\lambda_{n}) $  is the conditional pdf of ${\boldsymbol \lambda}_{N \backslash {n}}$ given in $\lambda_n$. Hence, the PEP can be re-written as
       \begin{equation}
  \label{PEP}
  \small
  \begin{aligned} 
\text{Pr} \{\mathbf{w} \to \mathbf{\hat{w}} \} &=  \int\limits_{{\boldsymbol \lambda}} \text{Pr} \{\mathbf{w} \to \mathbf{\hat{w}} \vert \boldsymbol {\lambda}\}f_{\Lambda}({\boldsymbol \lambda}) d{\boldsymbol \lambda} \\
 & =\prod_{k=1}^{K} \int\limits_{ {\boldsymbol \lambda}_{N \backslash {n}} }   \int \limits_{{\lambda}_n }\text{Pr} \{\mathbf{w} \to \mathbf{\hat{w}} \vert  \lambda_n,    n=\text{Ind}_k\}\\
 & \quad \quad  \times  f_{{{\Lambda}}_{n} \vert\lambda_{N \backslash {n}}}({ {\boldsymbol \lambda}_{N \backslash {n}}}\vert\lambda_{n}) f_{{ \lambda}_{n}}(\lambda_{n})d{\boldsymbol\lambda}_{N \backslash n} d{ \lambda}_n.
  \end{aligned}
\end{equation}
 
When the channel experiences Rayleigh fadings,  the conditional pdf  $ f_{{\Lambda}_{N \backslash n} \vert\lambda_{n}}({{\boldsymbol\lambda}_{N\backslash n }}\vert\lambda_{n}) $ can be approximated as a  Gaussian distribution for large $N$ with mean $\eta_{\boldsymbol{\lambda}_{N\backslash n } \mid \lambda_{n}}$
and covariance $\mathbf{C}_{\boldsymbol{\lambda}_{N\backslash n } \mid \lambda_{n}}$ expressed as \cite{Armstrong}
   \begin{equation}
   \label{eqN}
\begin{aligned}
\small
\boldsymbol{\eta}_{\boldsymbol{\lambda}_{N\backslash n } \mid \lambda_{n}} &=\lambda_{n} c_{\lambda_{n} \lambda_{n}}^{-1} \mathbf{c}_{\boldsymbol{\lambda}_{N\backslash n } \lambda_{n}} , \\
\mathbf{C}_{\boldsymbol{\lambda}_{N\backslash n } \mid \lambda_{n}} &=\mathbf{C}_{\boldsymbol{\lambda}_{N\backslash n } \boldsymbol{\lambda}_{N\backslash n }}-c_{\lambda_{n} \lambda_{n}}^{-1} \mathbf{c}_{\boldsymbol{\lambda}_{N\backslash n } \lambda_{n}} \mathbf{c}_{\boldsymbol{\lambda}_{N\backslash n } \lambda_{n}}^{\mathrm{H}},
\end{aligned}
\end{equation}
where $c_{\lambda_{m} \lambda_{n}}=E\left\{\lambda_{m} \lambda_{n}^{*}\right\}, \mathbf{c}_{\boldsymbol{\lambda}_{N\backslash n } \lambda_{n}}=[ c_{\lambda_{1}\lambda_{n}},\ldots,c_{\lambda_{n-1}\lambda_{n}}, $  $  c_{\lambda_{n+1}\lambda_{n}}, \ldots c_{\lambda_{N} \lambda_{n}}
]^{\mathcal {T}}$, and $\mathbf{C}_{\boldsymbol{\lambda}_{N\backslash n } \boldsymbol{\lambda}_{N\backslash n }}$ is the $(N-1)$-dimensional square matrix obtained by   removing  the $n$th row and column of  the frequency-domain channel covariance matrix  $ \mathbf{C}_{\boldsymbol{\lambda} \boldsymbol{\lambda}}=E\left\{\boldsymbol{\lambda} \boldsymbol{\lambda}^{\mathcal {H}}\right\}$.
    Here, we define the conditional random variable $t_{k}=z_{k} \mid \lambda_{n}$, and from (\ref{rk}), we have
    \begin{equation}
\small
\begin{aligned}
t_{k} = \phi  _{n,n}\lambda_{{n}} w_{k}+ \underbrace{\sum_{ m\in \xi_{N \backslash {n}} }  \phi  _{{n},m} \xi_{m} s_{m}}_{\text{Conditional ICI}_k}  +v_{{n}},  
\end{aligned}
\end{equation}
where the conditional random variable   $\xi _{m}=\lambda _{m}\mid \lambda _{n}$ is Gaussian with its mean  and variance  given by  $\eta _{m}= \lambda_{n} c_{\lambda_{n} \lambda_{n}}^{-1} \mathbf{c}_{{\lambda_{m}} \lambda_{n}}$ and $\sigma _{\xi _{m}}^{2} =\left[\mathbf{C}_{\boldsymbol{\lambda}_{N\backslash n } \mid \lambda_{n}}\right]_{m-1, m-1}, m=1,\cdots,N,m\ne n$, respectively. Consequently, by defining the zero-mean random variable $\zeta_{m} =\xi_{m}-\eta _{m} $, we obtain
\begin{equation}
\small
\label{eqt}
t_{k}=\phi_{n,n} \lambda_{n} w_{k}+\alpha_{n} \lambda_{n}+\beta_{n}+v_{n},
\end{equation}
where by means of (\ref{eqN}), $\alpha_{n}$ and $\beta_{n}$ are expressed as follows
\begin{equation}
\small
\label{pk}
\begin{aligned}
\alpha_{n}&=c_{\lambda_{n} \lambda_{n}}^{-1} \sum_{m\in \xi_{N \backslash {n}}} \phi_{n, m} c_{\lambda_{m} \lambda_{n}} s_{m}, \\
\beta_{n}&=\sum_{m\in \xi_{N \backslash {n}}} \phi_{n, m} \zeta_{m} s_{n}.
\end{aligned}
\end{equation}
It is obvious that $\alpha_{n}$ and $\beta_{n}$ do not depend on the specific value of $\lambda_{n}$ that characterizes the channel realization, but on the statistical characterization of the channel in the frequency domain. As can be seen from (\ref{eqt})-(\ref{pk}), the ICI is divided into two parts, $\lambda_{n}\alpha_{n}$ is proportional to the channel amplitude $\lambda_{n}$ of the useful signal, while the second part $\beta_{n}$ is independent of $\lambda_{n}$. The power of $\lambda_{n}\alpha_{n}$ and $\beta_{n}$ can be expressed as 
\begin{equation}
\label{eqdelta}
\small
\begin{aligned}
\sigma_{\alpha_{n}}^{2}&=\left|c_{\lambda_{n} \lambda_{n}}^{-1} \right|^{2} \sum_{m\in \xi_{N \backslash {n}}}\left|\phi_{n,m} c_{\lambda_{m} \lambda_{n}}\right|^{2}, \\
\sigma_{\beta_{n}}^{2}&=\sum_{m\in \xi_{N \backslash {n}}}\left|\phi_{n, m}\right|^{2}\left[C_{\boldsymbol{\lambda}_{N\backslash n } \mid \lambda_{n}}\right]_{m-1, m-1}.
\end{aligned}
 \end{equation}
    
We now turn to the evaluation of  $\text{Pr} \{\mathbf{w} \to \mathbf{\hat{w}} \}$ in  (\ref{PEP}). The conditioned PEP    is given by  
 \begin{equation}
 \label{PEP2}
 \small
 \begin{aligned}
    & \text{Pr}\left\{\mathbf{w} \rightarrow \mathbf{\hat{w}} \mid \lambda_{n}, n=\text{Ind}_k \right\}  \\
     & =\text{Pr}\left \{ \sum_{k=1}^{K} \left | z_{k}-\phi_{n,n}\lambda_{n}\hat{w }_{k}\right |^{2} 
\le \sum_{k=1}^{K} \left | z_{k}-\phi_{n,n}\lambda_{n}w_{k}\right |^{2}  \right \} \\
&=\text{Pr}\Bigg \{\sum_{k=1}^{K} \Big( 
\phi_{n, n}^{2} \lambda_{n}^{2} \Delta_{{k}}^{2}  + 2 \phi_{n, n} \lambda_{n}\Delta_{{k}} \left( \text{ICI}_k +v_{n}\right) \leq 0
 \Big)\Bigg\},
 \end{aligned}
\end{equation}
where $\Delta_{{k}} = w_{k} -\hat w_{k}$. To proceed, let us define
\begin{equation}
\small
\chi= \sum_{k=1}^{K}  \phi_{n, n} \lambda_{n}\Delta_{{k}}\left( \text{ICI}_k +v_{n}\right).
\end{equation}
  From (\ref{eqN})-(\ref{eqdelta}),  it is shown that the ICI at $k$th SCMA RE can be modeled as  Gaussian random variable  with     
 variance $\left|\lambda_{n}\right|^{2} \sigma_{ \alpha_{n}}^{2}+\sigma_{ \beta_{n}}^{2}$ based on central limit theorem. Hence, $\chi$ is also a Gaussian random variable with its variance given by 
\begin{equation}
\small
\sigma_{\chi}^{2}= \sum_{k=1}^{K}  \phi_{n , n}^{2} \lambda_{n}^{2}\Delta_{{k}}^2\left(\left|\lambda_{n}\right|^{2} \sigma_{ \alpha_{n}}^{2}+\sigma_{ \beta_{n}}^{2}+\sigma_{v}^{2}\right).
\end{equation}
Let $A=\frac{1}{2} \sum_{k=1}^{K}  \phi_{n, n}^{2} \lambda_{n}^{2}\Delta_{{k}}^2$, then (\ref{PEP2}) can be reformulated as
 \begin{equation}
 \small
 \label{eqcpep}
 \begin{aligned}
 &\text{Pr}\left\{\mathbf{w} \rightarrow \mathbf{\hat{w}} \mid \lambda_{n}, n=\text{Ind}_k \right\}  =\text{Pr}(\chi \geq A)=Q\left(A / \sigma_{\chi}\right)  \\
 &  = Q\left( \frac{  \sum_{k=1}^{K} \phi_{n, n}^{2} \lambda_{n}^{2} \Delta_{{k}}^2}{ 2\sqrt{\sum_{k=1}^{K}  \phi_{n , n}^{2} \lambda_{n}^{2}\Delta_{{k}}^2\left(\left|\lambda_{n}\right|^{2} \sigma_{ \alpha_{n}}^{2}+\sigma_{ \beta_{n}}^{2}+\sigma_{v}^{2}\right) }}\right) \\
 & \overset{\mathrm{i}}{\approx}   Q\left( \sqrt{\frac{1}{4} \sum_{k=1}^{K}  \gamma\left(\lambda_{n}\right) \Delta_{{k}}^2}\right),
 \end{aligned}
\end{equation}
where  $Q(x)=\frac{1}{\sqrt{2\pi } } \int_{x}^{\infty } e^{-\frac{t^{2} }{2} } dt$ is the $Q$-function, and
\begin{equation}
\small 
\gamma\left(\lambda_{n}\right)=\frac{\left|\lambda_{n}\right|^{2}\left|\phi_{n,n}\right|^{2}}{\left|\lambda_{n}\right|^{2} \sigma_{ \alpha_{n}}^{2}+\sigma_{\beta_{n}}^{2}+\sigma_{0}^{2}}, 
\end{equation}
is the conditional SINR. Step $\mathrm{i}$ is obtained based on the fact that the term $\left|\lambda_{n}\right|^{2} \sigma_{ \alpha_{n}}^{2}+\sigma_{\beta_{n}}^{2}+\sigma_{0}^{2}$ is approximately equal for each RE. 
It is worth noting that, when the  $N$ SCs fade simultaneously within an OFDM symbol duration, we have $\mathbf{c}_{\boldsymbol{\lambda}_{N\backslash n } \lambda_{n}} = \mathbf 0_{N-1 \times 1}$ and $\sigma_{\alpha_{n}}^{2}=0$. However, the condition of independent fading   is not realistic for all the SCs as it is not compatible with the usual hypothesis that the CP length is shorter than the OFDM symbol.  
  By applying the approximation $Q(x)\simeq \frac {1}{12}\exp (-x^{2}/2)+\frac {1}{4}\exp (-2x^{2}/3)$  \cite{SparseLiu}, (\ref{eqcpep}) can be approximated as  
 \begin{equation}
 \small
 \begin{aligned}
 \label{qAPPRO}
 & \text{Pr}\left\{\mathbf{w} \rightarrow \mathbf{\hat{w}} \mid \lambda_{n},  n=\text{Ind}_k \right\} \\
 & \simeq \frac{1}{12}\exp \left(-\frac{\sum_{k=1}^{K} \gamma\left(\lambda_{n}\right) \Delta_{{k}}^2}{8} \right) \\
 & \quad \quad \quad \quad \quad \quad + \frac{1}{4}\exp \left(-\frac{\sum_{k=1}^{K} \gamma\left(\lambda_{n}\right) \Delta_{{k}}^2}{6} \right).
 \end{aligned}
\end{equation}
Then, the unconditional PEP can be obtained by averaging over the channel statistics:  
  \begin{equation}
 \label{unPEP}
  \small
 \begin{aligned}
 &\text{Pr}\left\{\mathbf{w} \rightarrow \mathbf{\hat{w}} \right\}  \simeq \int_{0}^{\infty} \left( \frac{1}{12}\exp \left(-\frac{\sum_{k=1}^{K} \gamma\left(\lambda_{n}\right) \Delta_{{k}}^2}{8} \right) \right.\\& 
\left.+ \frac{1}{4}\exp \left(-\frac{\sum_{k=1}^{K} 
 \gamma \left(\lambda_{n}\right) \Delta_{{k}}^2}{6} \right) \right)
 \prod_{k=1}^{K} f_{\lambda_n}\left(\lambda_{n}\right)d\lambda_{n}.
 \end{aligned}
\end{equation}
For Rayleigh fading channels, we have $f_{\lambda_n}(\lambda_{n})=  \frac{2\left|\lambda_{n}\right|}{c_{\lambda_{n} \lambda_{n}}} \exp \left({-\frac{\left|\lambda_{n}\right|^{2}}{c_{\lambda_{n} \lambda_{n}}}} \right)$. In this case, (\ref{unPEP}) can be estimated by Monte Carlo methods or be expressed as the series expansion \cite{gradshteyn2014table}
\begin{equation}
\label{FPEP}
\small
\begin{aligned}
\text{Pr}\left\{\mathbf{w} \rightarrow \mathbf{\hat{w}} \right\}&\le
\frac{1}{12} \prod_{k=1}^{K} \left(1-\frac{1}{8 \upsilon } e^{\frac{4-\mu }{8 \upsilon }} \sum_{m=0}^{\infty} \frac{1}{m !}\left(\frac{1}{8 \upsilon }\right)^{m}\right.
\\& \quad \quad \left.\left(\frac{\mu }{\upsilon }\right)^{\frac{m}{2}} \mu ^{\frac{m+2}{2}} W_{\frac{-2-m}{2}, \frac{m+1}{2}}\left(\frac{1}{\upsilon }\right)\right)\\&
+ \frac{1}{4} \prod_{k=1}^{K}  \left(1-\frac{1}{6 \upsilon } e^{\frac{3-\mu }{6 \upsilon }} \sum_{m=0}^{\infty} \frac{1}{m !}\left(\frac{1}{6 \upsilon }\right)^{m}\right.
\\&  \quad \quad  \left.\left(\frac{\mu }{\upsilon }\right)^{\frac{m}{2}} \mu ^{\frac{m+2}{2}} W_{\frac{-2-m}{2}, \frac{m+1}{2}}\left(\frac{1}{\upsilon }\right)\right),
\end{aligned}
\end{equation}
with $\mu$ and $\upsilon$ given by
\begin{equation}
\small
\begin{aligned}
\mu  =\frac{c_{\lambda_{n}\lambda_{n}} \phi_{n,n}^{2}   \Delta_{{k}}^2}{\sigma_{\beta_{n}}^{2}+\sigma_{0}^{2}},
\upsilon  =\frac{c_{\lambda_{n} \lambda_{n}} \sigma_{\alpha_{n}}^{2}}{\sigma_{\beta_{n}}^{2}+\sigma_{0}^{2}},
\end{aligned}
\end{equation}
 respectively, and  $W(a,b,z)=e^{-z/2} z^{b+1/2} U(b-z+\frac{1}{2},1+2b,z )$, $U(c,d,x)=\frac{1}{\Gamma (c)} \int\limits_{0}^{1} \frac{t^{c-1}}{(1-t)^{d}} e^{\frac{-xt}{1-t} } dt$, and $\Gamma (\cdot )$ is the Gamma function.   

Following a similar derivation, we can obtain the PEP for   Gaussian channels, which  has the same form as (\ref{qAPPRO}), but with different expression of $ \gamma\left(\lambda_{n}\right)$ given by
 \begin{equation}
\small 
\overline{\gamma}\left(\lambda_{k}\right)=\frac{1 }{  \frac{J}{K}\left(1-{\sin^{2}\pi\varepsilon\over N^{2}\sin^{2}{\pi\varepsilon\over N}}\right)+\sigma_{0}^{2}}.
\end{equation}
Finally, by substituting $\text{Pr}\left\{\mathbf{w} \rightarrow \mathbf{\hat{w}} \right\}$ into (\ref{ABER}), we obtain  the BER of the OFDM-SCMA with CFO under both  Gaussian and multipath Rayleigh fading channels.  
 
\section{Numerical Results}
 In this section, numerical results are presented to evaluate the BER performance for the  CFO affected SCMA-OFDM systems.  Specifically, we consider a frequency-selective fading channel model consisting of $P=8$ paths, each with an independent Rayleigh distribution with delay and power profiles given by $[1,2,4,6,9,11,15,20]$ and $[0.36,    0.24,    0.15,    0.10,    0.06,    0.04,    0.025,   0.017],$ respectively.  Such model is suitable for simulating the propagation path responses in a scattering environment. 
 The  SCMA indicator matrix  that presented in Fig. \ref{facotGraph} with $J=6$, $K=4$ and $V=2$ is considered.  We employ the codebook proposed in \cite{huawei} with $M=4$ and a message passing algorithm detector for the  ML  performance \cite{oNOMA}. The   number of OFDM subcarriers and the length of CP are $N = 1024$ and $N_{cp}=32$, respectively. 

\begin{figure*}
\minipage{0.32\textwidth}
  \includegraphics[width=1\textwidth]{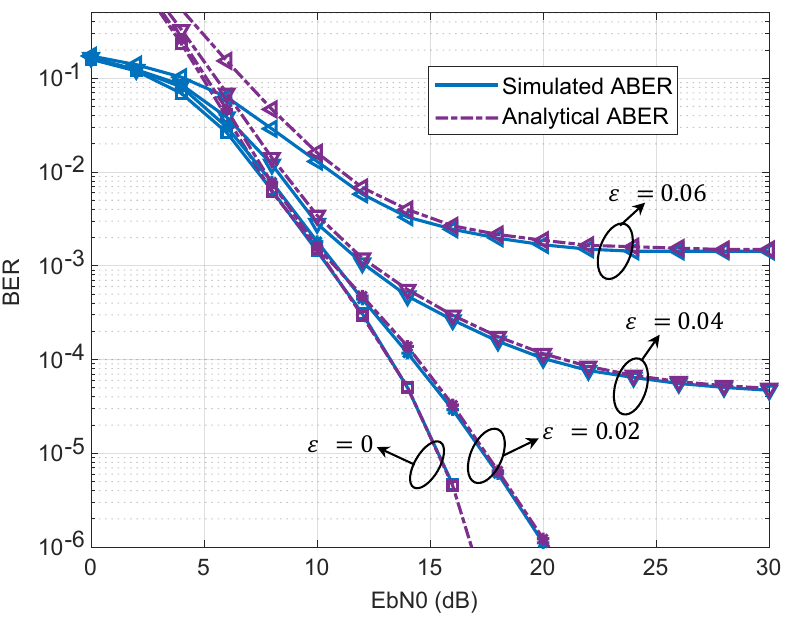}
\caption{BER performance of SCMA-OFDM over Gaussian channels.  }
\label{Gau}
\endminipage\hfill
\minipage{0.32\textwidth}
\includegraphics[width=1 \textwidth]{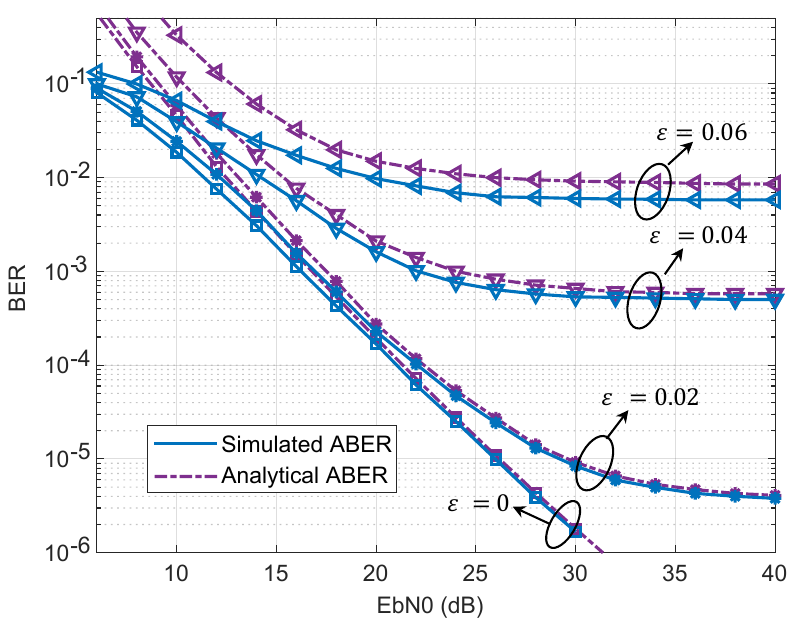}
 \caption{BER performance of SCMA-OFDM over multipath Rayleigh fading channels. }
\label{Ray}
\endminipage\hfill
\minipage{0.3\textwidth}%
\includegraphics[width=1.0\textwidth]{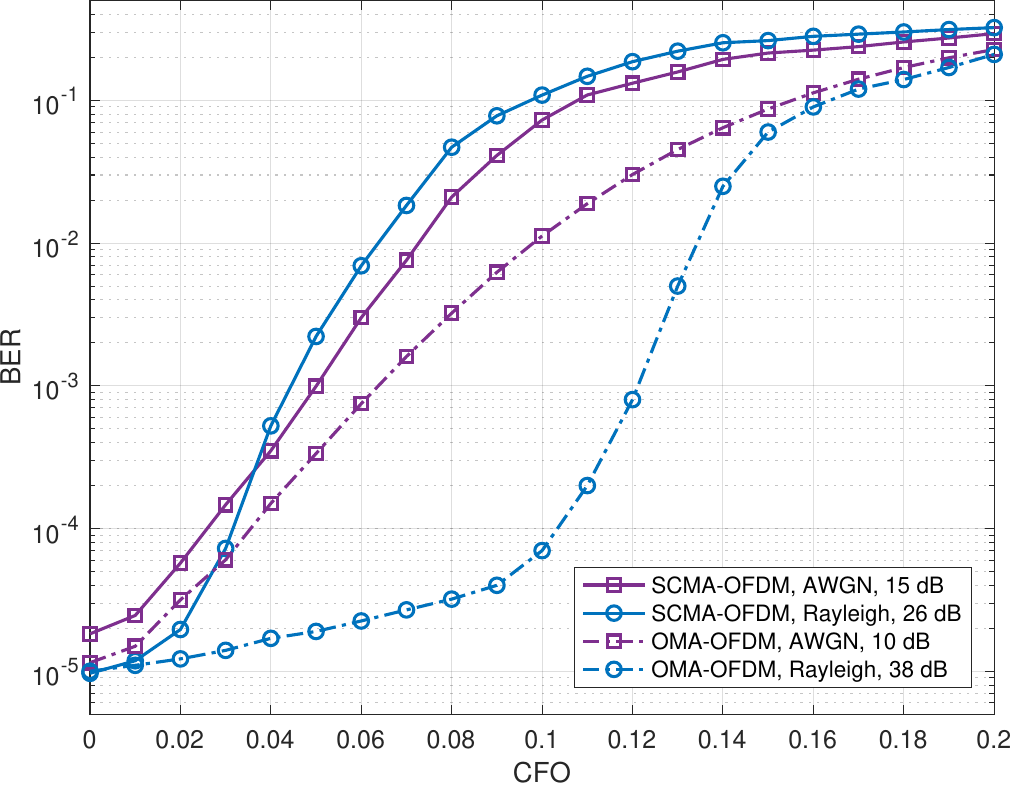}
 \caption{Simulated BER performance of SCMA-OFDM and OMA-OFDM with different CFO values. }
\label{CFO2}
\endminipage
\end{figure*}

Fig. \ref{Gau} and Fig. \ref{Ray}  depict  the BER performance of SCMA-OFDM in Gaussian and multipath Rayleigh fading channels, respectively.  Notably, all the simulated BERs   match  well with the theoretical analysis at medium-to-high SNR values for various normalized CFOs for $\varepsilon \leq 0.04$.  However, at low SNR values or large $\varepsilon $, a  discrepancy between the approximated and simulated BER is observed. 
This is due to the fact that the union bound in (\ref{ABER}) is not tight in low SNRs.  In addition, it is also observed that the BER performances of SCMA-OFDM systems  deteriorate significantly for CFO larger than $0.02$ in both Gaussian and multipath Rayleigh fading channels.

Fig. \ref{CFO2} shows the simulated BER performance of SCMA-OFDM and the OFDM with orthogonal multiple access (OMA-OFDM) with different CFO values in Gaussian and multipath Rayleigh fading channels. Specifically, the QPSK constellation with a maximum likelihood decoder for  a single user OMA-OFDM system  is considered. Obviously, SCMA-OFDM is more sensitive to CFO than that of the OMA-OFDM systems, especially in  Rayleigh fading channels.  
 It is also observed that the SCMA-OFDM systems in Rayleigh fading channels are slightly more sensitive to CFO than in Gaussian channels.  The BER performance  of SCMA-OFDM system   deteriorates shapely for CFOs larger than $0.02$,  whereas the OMA-OFDM  deteriorates significantly for CFOs larger than $0.1$ over Rayleigh fading channels.

\vspace{-1em}
\section{Conclusion}
In this paper, we have investigated the impact of CFO impairments on the performance of SCMA-OFDM systems. Although SCMA-OFDM offers high spectral efficiency and enables massive connectivity, it is sensitive to the effects of CFO. We have observed that the BER performance degrades significantly when the normalized CFO exceeds $0.02$. Therefore, the impact of CFO should be taken into account when designing SCMA-OFDM systems. As a future work, other existing modulation schemes that are robust to CFO effects, such as orthogonal time frequency space, can be explored for the SCMA systems.

\vspace{-1em}

\bibliography{ref} 
\bibliographystyle{IEEEtran}

\end{document}